\documentclass[12pt]{iopart}

\usepackage{graphicx,bm}

\begin{document}

\title{Comment on ``The hard sphere quantum propagator: exact results via partial wave analysis''}

\author{Arseni Goussev$^1$, Orestis Georgiou$^2$, Valeriy Slastikov$^3$}
\address{$^1$ Department of Mathematics, Physics and Electrical Engineering, Northumbria University, Newcastle Upon Tyne NE1 8ST, UK}
\address{$^2$ Ultrahaptics, The West Wing, Glass Wharf, Bristol BS2 0EL, UK}
\address{$^3$ School of Mathematics, University of Bristol, University Walk, Bristol BS8 1TW, UK}

\begin{abstract}
	There is no known exact expression for the propagator of a non-relativistic particle colliding with a hard sphere. De Prunel\'e (2008 {\it J.~Phys.~A:~Math.~Theor.} {\bf 41} 255305) derived a partial wave expansion of the propagator and compared it against some known approximations, including the semiclassical Van Vleck-Gutzwiller (VG) propagator; the VG propagator was evaluated entirely numerically. Here we point out that the VG propagator for the particle-sphere problem admits an analytic expression in terms of elementary functions.
\end{abstract}


Exact closed-form expressions for quantum propagators in time-dependent scattering systems are very rare. Even such a `simple' scenario as a non-relativistic point particle colliding with a hard sphere has no known exact propagator. In \cite{prunele}, de Prunel\'e obtained an infinite-series partial wave expansion for the particle-sphere propagator and compared the accuracy of his expansion against the predictions of two available approximations: the approximation due to Cio and Berne \cite{cao-berne} and the semiclassical (short-wavelength) Van Vleck-Gutzwiller (VG) propagator \cite{gutzwiller}.
\begin{figure}[h!]
	\centering
	\includegraphics[width=0.4\textwidth]{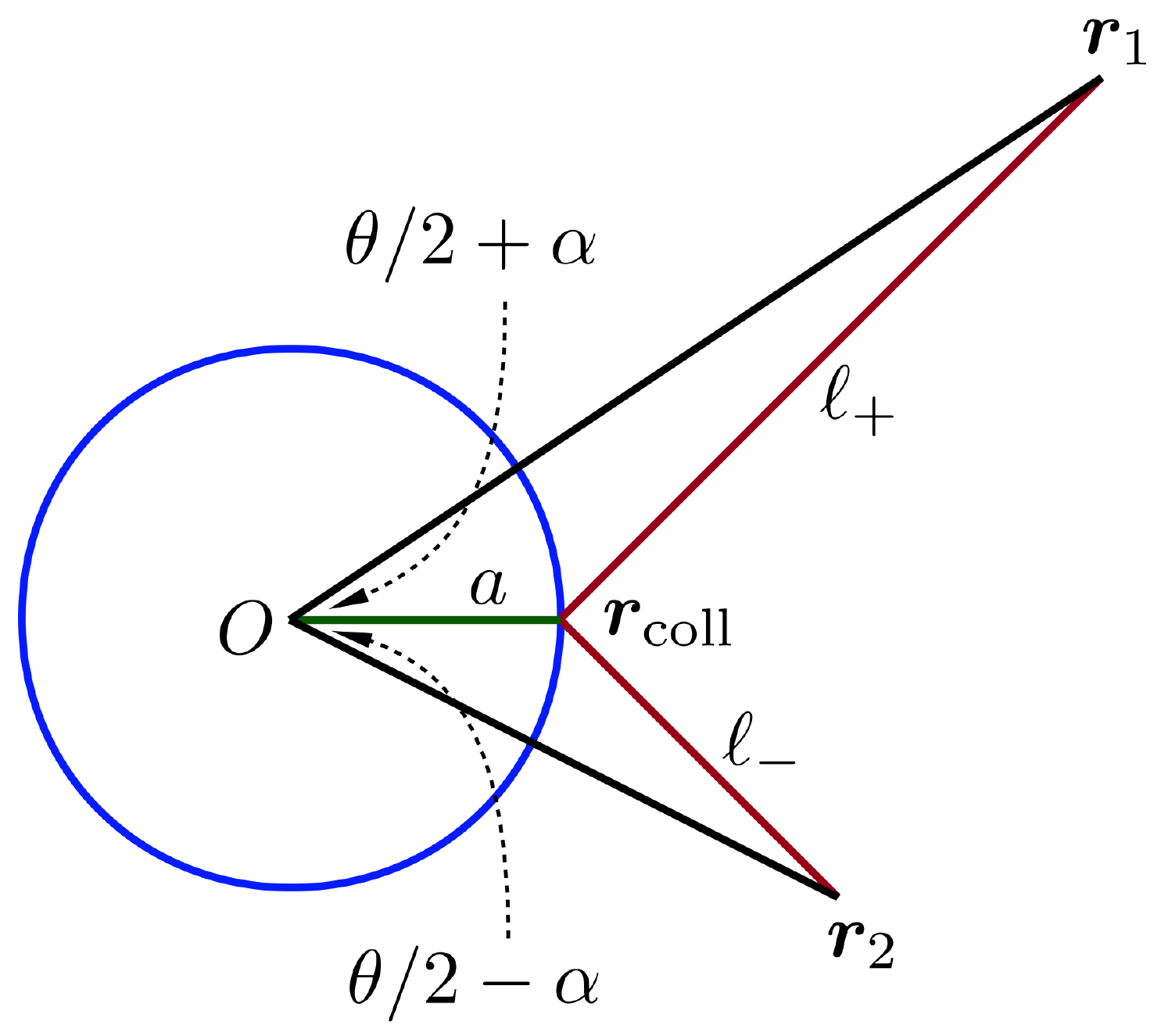}
	\caption{Collision path.}
	\label{fig}
\end{figure}
For the particle-sphere problem, the VG propagator takes the form (cf.~(24) in \cite{prunele})
\begin{equation}
	\langle \bm{r}_2 | U_{\mathrm{VG}} | \bm{r}_1 \rangle = \langle \bm{r}_2 | U_0 | \bm{r}_1 \rangle - \frac{\sqrt{|D|}}{(2 \pi i \hbar)^{3/2}} \exp \left( \frac{i S}{\hbar} \right) \,,
\label{VG}
\end{equation}
where $\langle \bm{r}_2 | U_0 | \bm{r}_1 \rangle$ is the free-particle propagator, $S$ is the classical action corresponding to the collision path leading from point $\bm{r}_1$ to $\bm{r}_2$ in time $t$ (see figure~\ref{fig}), and $D = \det \left( -\frac{\partial^2 S}{\partial \bm{r}_2 \partial \bm{r}_1} \right)$ is the Van Vleck determinant. It is assumed that the origin $O$ of the coordinate frame coincides with the sphere centre (figure~\ref{fig}) and that points $\bm{r}_1$ and $\bm{r}_2$ do not lie in the geometric shadow of one another. The action is given by
\begin{equation}
	S = \frac{M}{2 t} L^2 \,, \qquad L = \ell_+ + \ell_- \,,
\end{equation}
where $M$ is the particle mass, and $\ell_+$ (respectively $\ell_-$) is the distance between $\bm{r}_1$ (respectively $\bm{r}_2$) and the collision point $\bm{r}_{\mathrm{coll}}$, see figure~\ref{fig}. Path length $L$ is a function of the sphere radius $a$, lengths $r_1 \equiv |\bm{r}_1|$ and $r_2 \equiv |\bm{r}_2|$, and angle $\theta$ between $\bm{r}_1$ and $\bm{r}_2$. The absence of an explicit expression for $L(a,r_1,r_2,\theta)$ forced de Prunel\'e to compute action $S$ via numerical minimization and, subsequently, to evaluate matrix $\frac{\partial^2 S}{\partial \bm{r}_2 \partial \bm{r}_1}$ using finite differences. Below we show that function $L(a,r_1,r_2,\theta)$ is expressible in elementary functions.

Denoting the angle between $\bm{r}_1$ and $\bm{r}_{\mathrm{coll}}$ by $\theta/2 + \alpha$ (see figure~\ref{fig}), we have 
\begin{equation}
	\ell_\pm = \sqrt{a^2 + r_1 - 2 a r_1 \cos \left( \theta / 2 \pm \alpha \right)} \,.
\end{equation}
An exactly solvable equation for the unknown angle $\alpha$ can be obtained in the following three steps. Firstly, we notice that the cosine of the angle between $\bm{r}_{\mathrm{coll}}$ and $(\bm{r}_1 - \bm{r}_{\mathrm{coll}})$ is equal to the cosine of the angle between $\bm{r}_{\mathrm{coll}}$ and $(\bm{r}_2 - \bm{r}_{\mathrm{coll}})$. This leads to $\bm{r}_{\mathrm{coll}} \cdot (\bm{r}_1 - \bm{r}_{\mathrm{coll}}) / (a \ell_+) = \bm{r}_{\mathrm{coll}} \cdot (\bm{r}_2 - \bm{r}_{\mathrm{coll}}) / (a \ell_-)$, or
\begin{equation}
	\frac{r_1 \cos (\theta / 2 + \alpha) - a}{\ell_+} = \frac{r_2 \cos (\theta / 2 - \alpha) - a}{\ell_-} \,.
\label{eq1}
\end{equation}
Secondly, we use the fact that $\alpha$ must extremize the length of the reflection path, $L = \ell_+ + \ell_-$, i.e. $\partial L / \partial \alpha = 0$. This leads to
\begin{equation}
	\frac{r_1 \sin (\theta / 2 + \alpha)}{\ell_+} = \frac{r_2 \sin (\theta / 2 - \alpha)}{\ell_-} \,.
\label{eq2}
\end{equation}
Thirdly, dividing (\ref{eq1}) by (\ref{eq2}), and so eliminating $\ell_+$ and $\ell_-$, and performing straightforward trigonometric manipulations, we obtain
\begin{equation}
	\sin \alpha \cos \alpha - u \sin \alpha + v \cos \alpha = 0 \,,
\label{eq3}
\end{equation}
where
\begin{equation}
	u = \frac{a}{2} \left( \frac{1}{r_1} + \frac{1}{r_2} \right) \cos \frac{\theta}{2} \,, \qquad v = \frac{a}{2} \left( \frac{1}{r_1} - \frac{1}{r_2} \right) \sin \frac{\theta}{2} \,.
\end{equation}
Notice that $0 < u < 1$ and $-\frac{1}{2} < v < \frac{1}{2}$. The substitution
\begin{equation}
	x = \tan \frac{\alpha}{2}
\end{equation}
transforms equation (\ref{eq3}) into
\begin{equation}
	v x^4 + 2 (1 + u) x^3 - 2 (1 - u) x - v = 0 \,.
\label{eq4}
\end{equation}
This quartic equation, and consequently equation (\ref{eq3}), is exactly solvable (see, e.g., chapter 3 in \cite{abramowitz}). While an explicit expression for $\alpha$ as a function of $u$ and $v$ is too lengthy to be presented in this Comment, it can be readily obtained using such symbolic packages as Mathematica or Maple. Here we only report a simple (but extremely accurate) approximate solution obtained by regarding $v \in \left( -\frac{1}{2} , \frac{1}{2} \right)$ as a small parameter:
\begin{equation}
	\alpha = -\frac{v}{1 - u} - \frac{(1 + 2 u) v^3}{6 (1 - u)^4} + \mathcal{O}(v^5) \,.
\end{equation}
The exact solution to equation (\ref{eq3}), or the approximation given above, makes it possible to write an analytic expression for the VG propagator (\ref{VG}), thus avoiding the necessity of potentially expensive numerical computations.

\end{document}